\documentclass[a4paper, conference]{IEEEtran}

\usepackage{cite}
\usepackage{amsmath,amssymb,amsfonts}
\usepackage{algorithmic}
\usepackage{graphicx}
\usepackage{textcomp}
\usepackage{xcolor}
\usepackage{paralist}
\usepackage{eurosym}
\usepackage{flushend}
\usepackage{url}
\usepackage[utf8]{inputenc}
\usepackage[colorlinks=true]{hyperref}
\usepackage{soul}
\usepackage{caption}
\usepackage[list=true]{subcaption}
\usepackage{booktabs}
\usepackage{tabu}
\usepackage{wrapfig}
\usepackage{arydshln}
\usepackage[normalem]{ulem}
\usepackage{multirow}
\usepackage{siunitx}
\usepackage{adjustbox}
\usepackage{array}
\usepackage{diagbox}

\begin{document}

\title{Attention-Based Fusion of IQ and FFT Spectrograms with AoA Features for GNSS Jammer Localization}

\author{\IEEEauthorblockN{Lucas Heublein\IEEEauthorrefmark{1}, Christian Wielenberg\IEEEauthorrefmark{1}, Thorsten Nowak\IEEEauthorrefmark{2}, Tobias Feigl\IEEEauthorrefmark{1}, Christopher Mutschler\IEEEauthorrefmark{1},
    \underline{Felix Ott}\IEEEauthorrefmark{1}}
  \IEEEauthorblockA{\IEEEauthorrefmark{1}Fraunhofer Institute for Integrated Circuits IIS, 90411 Nürnberg, Germany}
  \IEEEauthorblockA{\IEEEauthorrefmark{2}Diehl Defence GmbH \& Co. KG, 90552 Röthenbach an der Pegnitz, Germany}
  \IEEEauthorblockA{\{lucas.heublein, christian.wielenberg, tobias.feigl, christopher.mutschler, felix.ott\}@iis.fraunhofer.de}
  \IEEEauthorblockA{thorsten.nowak@diehl-defence.com}
}

\IEEEoverridecommandlockouts
\IEEEpubid{\makebox[\columnwidth]{
979-8-3503-8078-1/24/\$31.00~\copyright2025
IEEE \hfill} \hspace{\columnsep}\makebox[\columnwidth]{ }}

\maketitle

\begin{abstract}
Jamming devices disrupt signals from the global navigation satellite system (GNSS) and pose a significant threat by compromising the reliability of accurate positioning. Consequently, the detection and localization of these interference signals are essential to achieve situational awareness, mitigating their impact, and implementing effective counter-measures. Classical Angle of Arrival (AoA) methods exhibit reduced accuracy in multipath environments due to signal reflections and scattering, leading to localization errors. Additionally, AoA-based techniques demand substantial computational resources for array signal processing. In this paper, we propose a novel approach for detecting and classifying interference while estimating the distance, azimuth, and elevation of jamming sources. Our benchmark study evaluates 128 vision encoder and time-series models to identify the highest-performing methods for each task. We introduce an attention-based fusion framework that integrates in-phase and quadrature (IQ) samples with Fast Fourier Transform (FFT)-computed spectrograms while incorporating 22 AoA features to enhance localization accuracy. Furthermore, we present a novel dataset of moving jamming devices recorded in an indoor environment with dynamic multipath conditions and demonstrate superior performance compared to state-of-the-art methods.
\end{abstract}
\IEEEpeerreviewmaketitle

\section{Introduction}
\label{label_introduction}

The localization accuracy of GNSS receivers is severely degraded by interference signals originating from jamming devices~\cite{ott_heublein_icl,raichur_heublein,heublein_raichur_ion,heublein_feigl_posnav,gaikwad_heublein,manjunath_heublein}. This issue has become increasingly pronounced in recent years due to the growing availability of low-cost jamming devices~\cite{merwe_franco}. Therefore, mitigating interference signals or neutralizing the source of interference is imperative. To implement effective counter-measures, it is essential to detect and classify interference signals and accurately determine the location of their source~\cite{raichur_ion_gnss,brieger_ion_gnss}. Classical jammer localization techniques encompass methods such as Received Signal Strength (RSS), Angle of Arrival (AoA)~\cite{schmidt,zhu_chen_yang,fuchs_gardill}, Direction of Arrival (DoA)~\cite{papageorgiou_sellathurai,feintuch_tabrikian}, Time Difference of Arrival (TDoA), and Frequency Difference of Arrival (FDoA)~\cite{qiao_lu_lin}. In AoA and DoA techniques, the direction from which the interference signal arrives at the receiver is estimated~\cite{yardibi_li_stoica}. However, these approaches are susceptible to multipath interference, wherein signals reflect off surfaces such as buildings, resulting in multiple signal paths reaching the receiver~\cite{heublein_feigl_crpa}. This phenomenon can introduce errors in angle estimation, leading to inaccuracies in localization. Moreover, achieving precise angle measurements often requires large antenna arrays and sophisticated hardware, which can be both costly and complex to implement.

Machine learning (ML) techniques have been introduced to overcome the limitations of traditional AoA and DoA methods in GNSS interference localization. ML algorithms are capable of modeling and predicting multipath effects by learning complex patterns from data~\cite{heublein_feigl_crpa}. These algorithms can correct distortions induced by multipath propagation, leading to more accurate angle estimations. Additionally, ML techniques can detect the presence of obstructions and adjust their estimations accordingly, thereby enhancing performance in non-line-of-sight (NLoS) conditions~\cite{feintuch_tabrikian}. Recent approaches aim to leverage the strengths of both traditional AoA techniques and ML methods~\cite{zeng_gong_liu,papageorgiou_sellathurai,nguyen_noubir,feintuch_tabrikian,zhu_chen_yang,fuchs_gardill,zheng_liu_sun,brieger_ion_gnss}. Our objective is to classify and localize interference sources under NLoS conditions, where multipath effects are prevalent. This will be achieved by optimally integrating IQ components with FFT-derived spectrograms, each analyzed through separate models, alongside classical AoA statistical features~\cite{wu_zhou_shen}. By fusing these elements, we aim to enhance the robustness and accuracy of interference source localization, even in challenging environments where traditional methods may struggle.

\begin{figure*}[!t]
    \centering
    \includegraphics[width=1.0\linewidth]{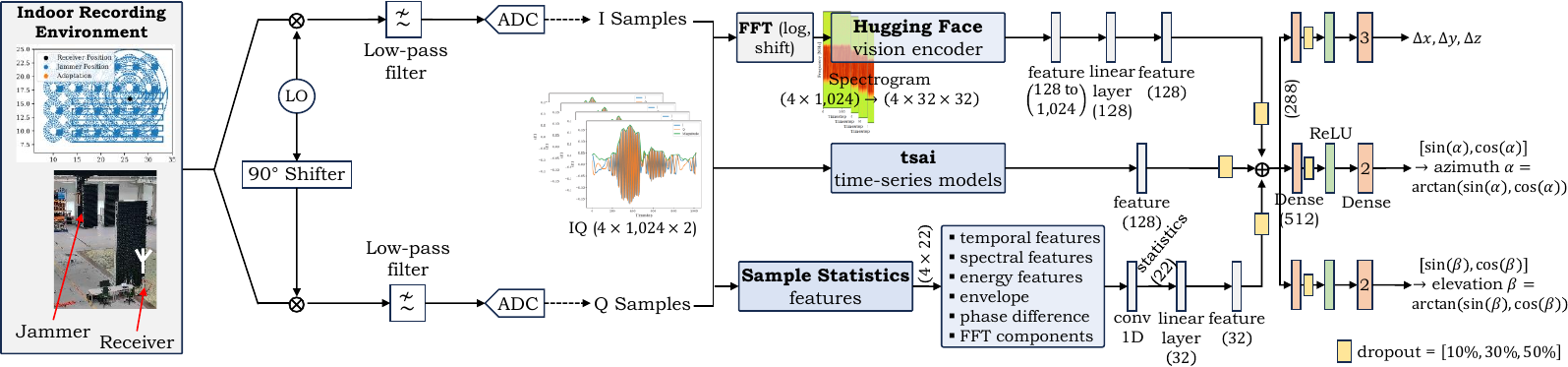}
    \caption{Overview of our pipeline. Raw measurements are collected from mobile jamming devices and processed into IQ samples. Vision encoders are trained with FFT-computed spectrograms, and combined with time-series models and statistical features. After concatenation, all features are passed through dense layers and ReLU activation for three distinct losses.}
    \label{figure_overview_method}
\end{figure*}

\textbf{Contributions.} The primary objective of this work is to classify and localize GNSS jamming devices by predicting their distance, azimuth, and elevation through the fusion of IQ data and FFT-computed spectrograms. To achieve this, we benchmark 110 vision encoder ML models and 18 novel time-series models. Additionally, we incorporate classical features from 22 traditional AoA techniques~\cite{wu_zhou_shen}, such as temporal, spectral, energy, envelope, and phase difference features, and compare the results against an FFT component fusion algorithm proposed by Zeng et al.~\cite{zeng_gong_liu}. We introduce a fusion methodology based on attention mechanisms and dropout to enhance model generalization. For comprehensive evaluation, we collect a large-scale dataset of moving jamming devices under dynamic environmental conditions, specifically accounting for varying multipath effects, within an indoor industrial hall.

\section{Related Work}
\label{label_related_work}

Qiao et al.~\cite{qiao_lu_lin} conducted a comprehensive survey on amplitude, phase, and spatial spectrum estimation techniques for interference source direction finding. Their study encompasses various methods, including AoA, DoA, TDoA, and FDoA, with an emphasis on theoretical foundations and classical approaches. Among spatial spectrum estimation techniques, the multiple signal classification (MUSIC) algorithm~\cite{schmidt} is recognized as one of the most prominent. \cite{papageorgiou_sellathurai} introduced a convolutional neural network (CNN)-based approach that predicts DoA using the sample covariance matrix estimate. Similarly, Nguyen et al.~\cite{nguyen_noubir} proposed a universal anti-jamming framework leveraging CNNs to detect the presence of interference, estimate the number of emissions, and determine their respective phases. Yardibi et al.~\cite{yardibi_li_stoica} developed a nonparametric, least squares-based iterative adaptive approach (IAA) for amplitude and phase estimation. Additionally, Feintuch et al.~\cite{feintuch_tabrikian} presented a CNN-based DoA estimation method for spatial spectrum analysis in multisource scenarios, where the number of sources is a priori unknown and the environment includes non-Gaussian interference. However, their evaluation was conducted using a simulated dataset.

In the domain of MIMO radar systems, Zhu et al.~\cite{zhu_chen_yang} introduced an AoA estimation algorithm based on Transformers (AAETR). This approach demonstrated superior computational efficiency compared to IAA while also exhibiting zero-shot simulation-to-real transferability for autonomous driving applications. Given the limitations associated with collecting labeled outdoor datasets and the capabilities of Transformer-based models, these factors motivate the collection of a labeled indoor dataset. However, it is important to note that AAETR is not suitable for jammer localization. Within the context of automotive radar, Fuchs et al.~\cite{fuchs_gardill} conducted an evaluation of model-based and data-driven approaches for angle estimation algorithms, emphasizing the benefits of a proposed hybrid data generation method. Additionally, Zheng et al.~\cite{zheng_liu_sun} reformulated the IAA algorithm as a series of gated recurrent unit (GRU) layers, which enhances generalization while mitigating the high computational costs associated with traditional implementations. Zeng et al.~\cite{zeng_gong_liu} introduced the multi-channel attentive feature fusion (McAFF) framework for radio frequency (RF) fingerprinting. This approach leverages multi-channel neural features extracted from various representations of RF signals, including IQ samples, carrier frequency offsets, FFT coefficients, and short-time Fourier Transform (STFT) coefficients. The method was evaluated using WiFi data for device identification. In this work, we adopt McAFF as a baseline for feature fusion and GNSS jammer localization.
\section{Methodology}
\label{label_method}

\begin{figure*}[!t]
    \centering
    \includegraphics[width=1.0\linewidth]{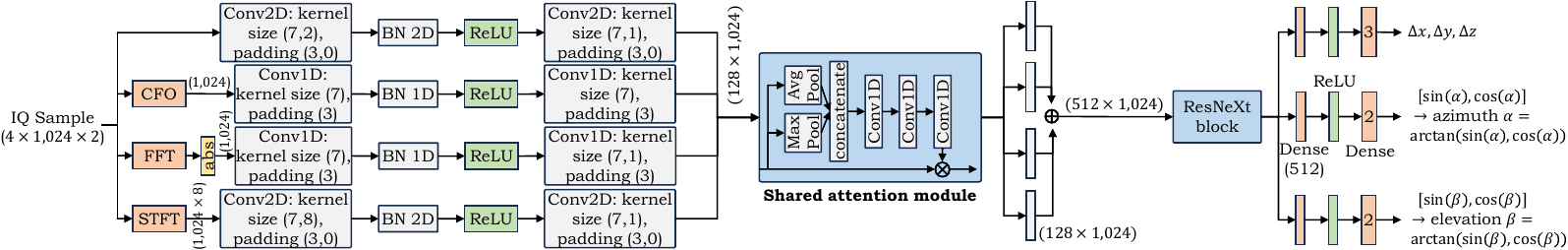}
    \caption{Overview of the adapted fusion baseline McAFF~\cite{zeng_gong_liu} for GNSS jammer localization.}
    \label{figure_fusion_baseline}
\end{figure*}

\textbf{Overview.} In this section, we provide an overview of our methodology (refer to Figure~\ref{figure_overview_method}). Initially, we collect a dataset of GNSS raw measurements within an indoor industrial environment equipped with a mobile jamming device, utilizing a 3D positioning system (see Section~\ref{label_experiments}). This dataset comprises raw measurements along with corresponding $x$-, $y$-, and $z$-coordinate labels relative to the receiver antenna. These data are subsequently processed into raw IQ samples for each antenna patch. We employ these IQ samples in three distinct processing paths: (1) We compute spectrograms of size $4 \times 1024$ using an FFT and reshape them to dimensions $4 \times 32 \times 32$. These spectrograms are then used to train a vision encoder from the Hugging Face model repository. The best-performing model is selected for feature fusion, followed by a linear layer. (2) The IQ samples are directly processed using 18 different time-series models from the tsai library. The most effective model is selected for fusion, producing a feature representation of size 128. (3) We compute 22 commonly used AoA estimation features for each antenna patch, following the framework proposed by Wu et al.~\cite{wu_zhou_shen}. These features are further processed using a one-dimensional convolutional layer with a kernel size of 1, followed by a linear layer, resulting in a feature size of 32. The extracted features from all three paths are concatenated into a single representation of size 288. For each regression task, we employ a dense layer with 512 units and ReLU activation, followed by a final dense layer of size 2 or 3, which predicts the relative displacement $[\Delta x, \Delta y, \Delta z]$ between the jamming device and the antenna, as well as the azimuth angle $\alpha$ and elevation angle $\beta$. We utilize \textit{tanh} activation before the computation of azimuth and elevation. To balance the loss functions effectively, we optimize the weighting factor $\gamma$ over the set $\{\frac{1}{1000}, \frac{1}{100}, \frac{1}{10}, \frac{1}{5}, 1, 10\}$ for the distance metric. To mitigate overfitting, we apply dropout rates of $\{10\%, 30\%, 50\%\}$ both before concatenation and after the final dense layer.

\textbf{Fusion Baseline.} We adopt the architecture from McAFF~\cite{zeng_gong_liu} as the state-of-the-art baseline for our approach to GNSS jammer localization. Figure~\ref{figure_fusion_baseline} provides an overview of the processing pipeline. The input IQ sample is processed through four distinct paths: (1) raw measurements undergo processing using 2D convolutional layers, (2) the accumulated carrier frequency offset (CFO) captures phase differences, (3) FFT coefficients serve as a frequency-domain representation of the signal, and (4) short-time Fourier Transform (STFT) coefficients, which represent a sequence of discrete Fourier Transforms, are processed similarly. These features are then integrated through a shared attention module, after which their outputs are concatenated. Subsequently, a ResNeXt~\cite{xie_girshick} block extracts relevant features, and the final predictions for distance, azimuth, and elevation are obtained using the same loss function as in our proposed method.
\section{Experiments}
\label{label_experiments}

\newcommand\ab{0.19}
\begin{figure}[!t]
    \centering
	\begin{minipage}[t]{\ab\linewidth}
        \centering
        \includegraphics[trim=1700 0 65 0, clip, width=1.0\linewidth]{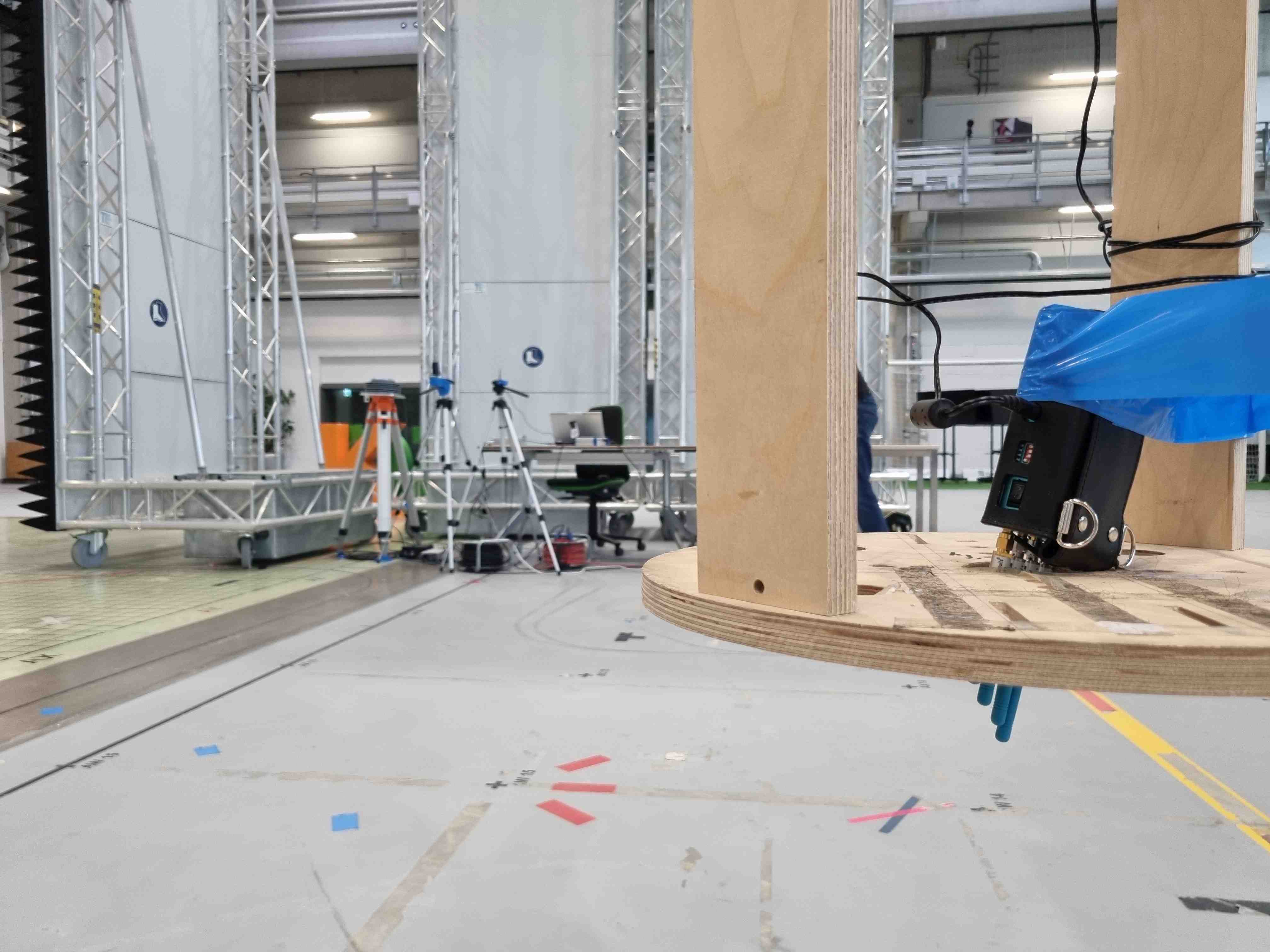}
        \subcaption{Jammer.}
        \label{figure_dataset_pictures1}
    \end{minipage}
    \hfill
	\begin{minipage}[t]{\ab\linewidth}
        \centering
        \includegraphics[trim=0 0 0 0, clip, width=1.0\linewidth]{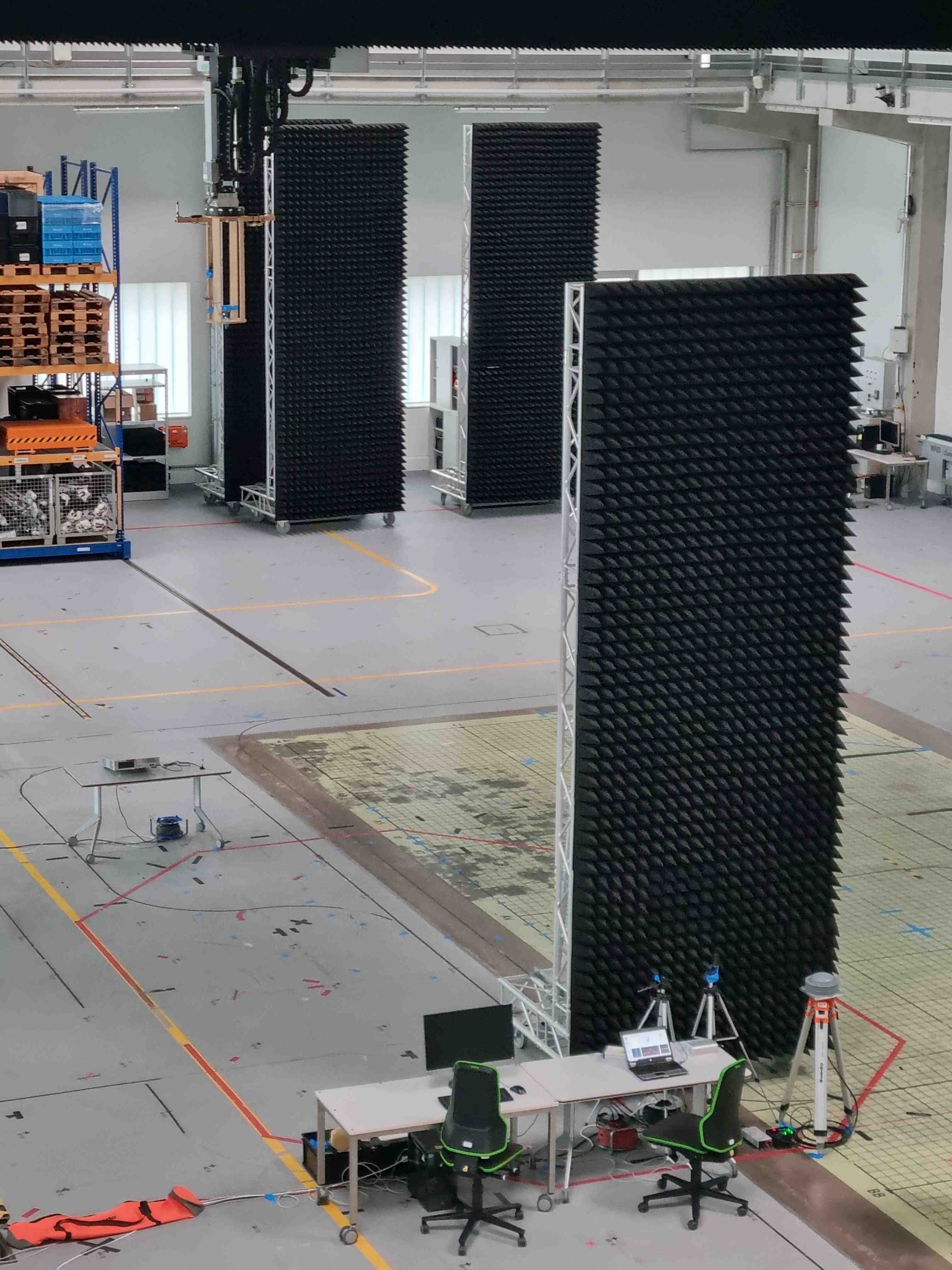}
        \subcaption{Wall 1.}
        \label{figure_dataset_pictures2}
    \end{minipage}
    \hfill
	\begin{minipage}[t]{\ab\linewidth}
        \centering
        \includegraphics[trim=0 0 0 0, clip, width=1.0\linewidth]{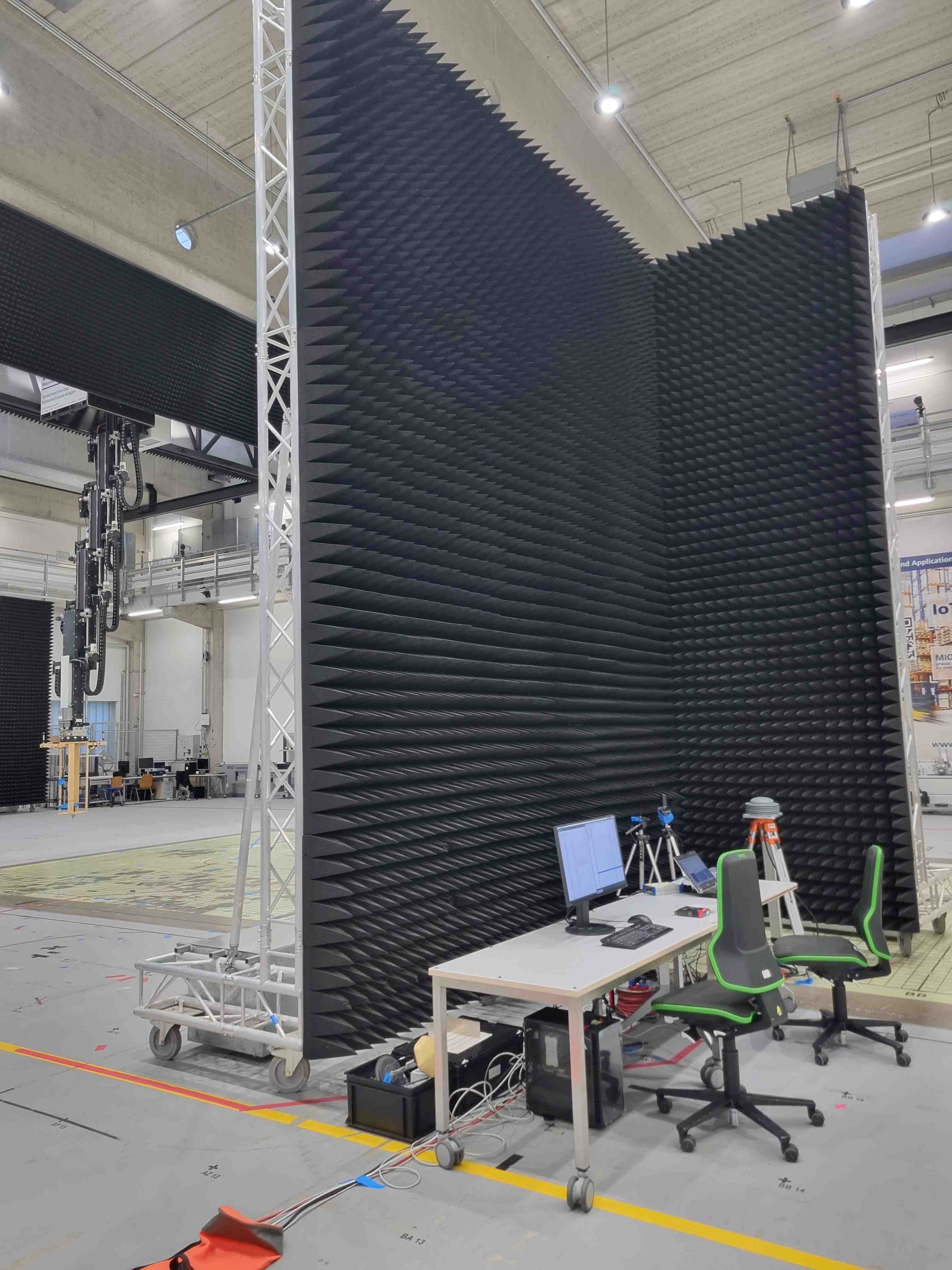}
        \subcaption{Wall 2.}
        \label{figure_dataset_pictures3}
    \end{minipage}
    \hfill
	\begin{minipage}[t]{\ab\linewidth}
        \centering
        \includegraphics[trim=0 0 0 0, clip, width=1.0\linewidth]{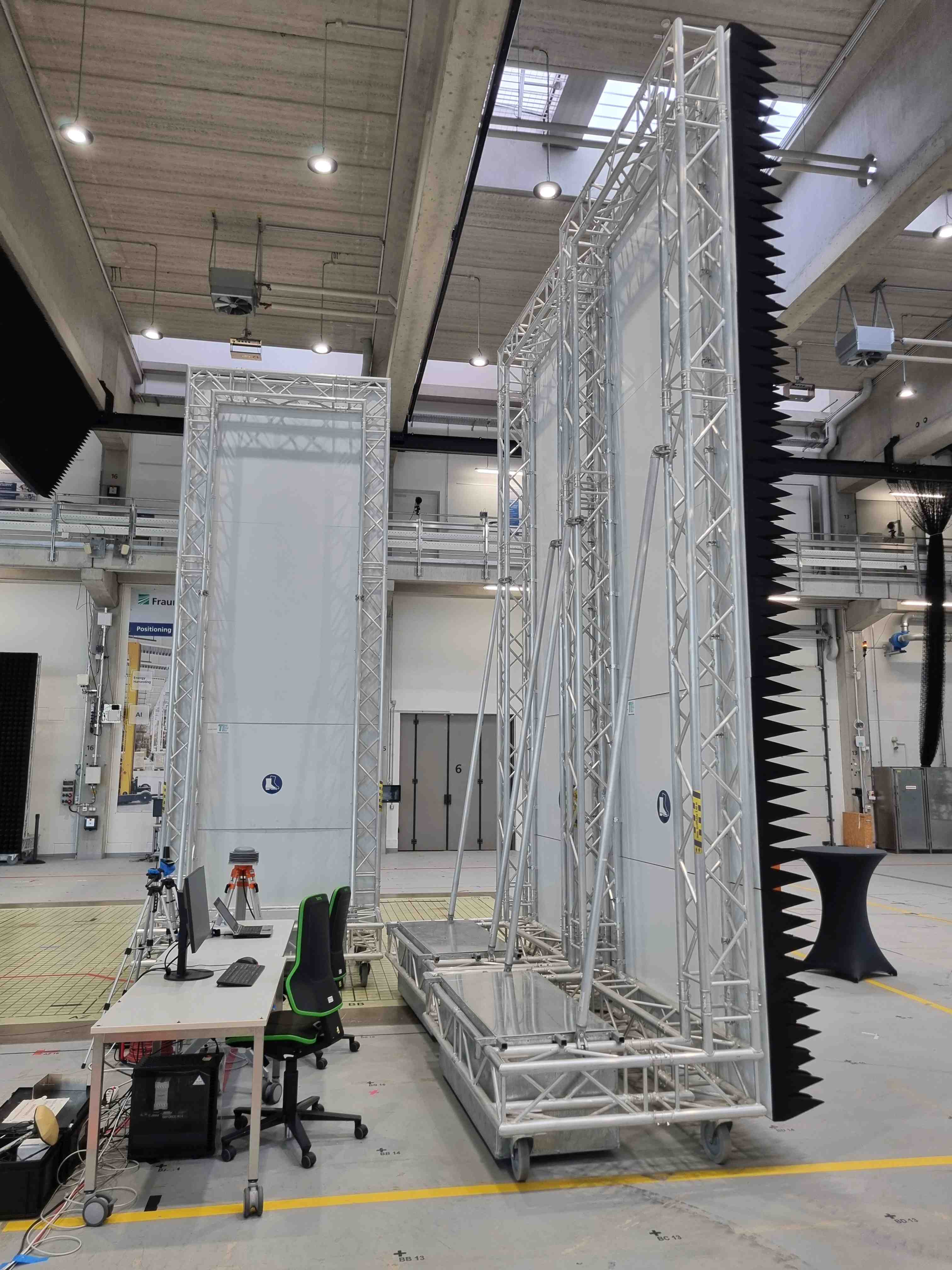}
        \subcaption{Wall 3.}
        \label{figure_dataset_pictures4}
    \end{minipage}
    \hfill
	\begin{minipage}[t]{\ab\linewidth}
        \centering
        \includegraphics[trim=0 0 0 0, clip, width=1.0\linewidth]{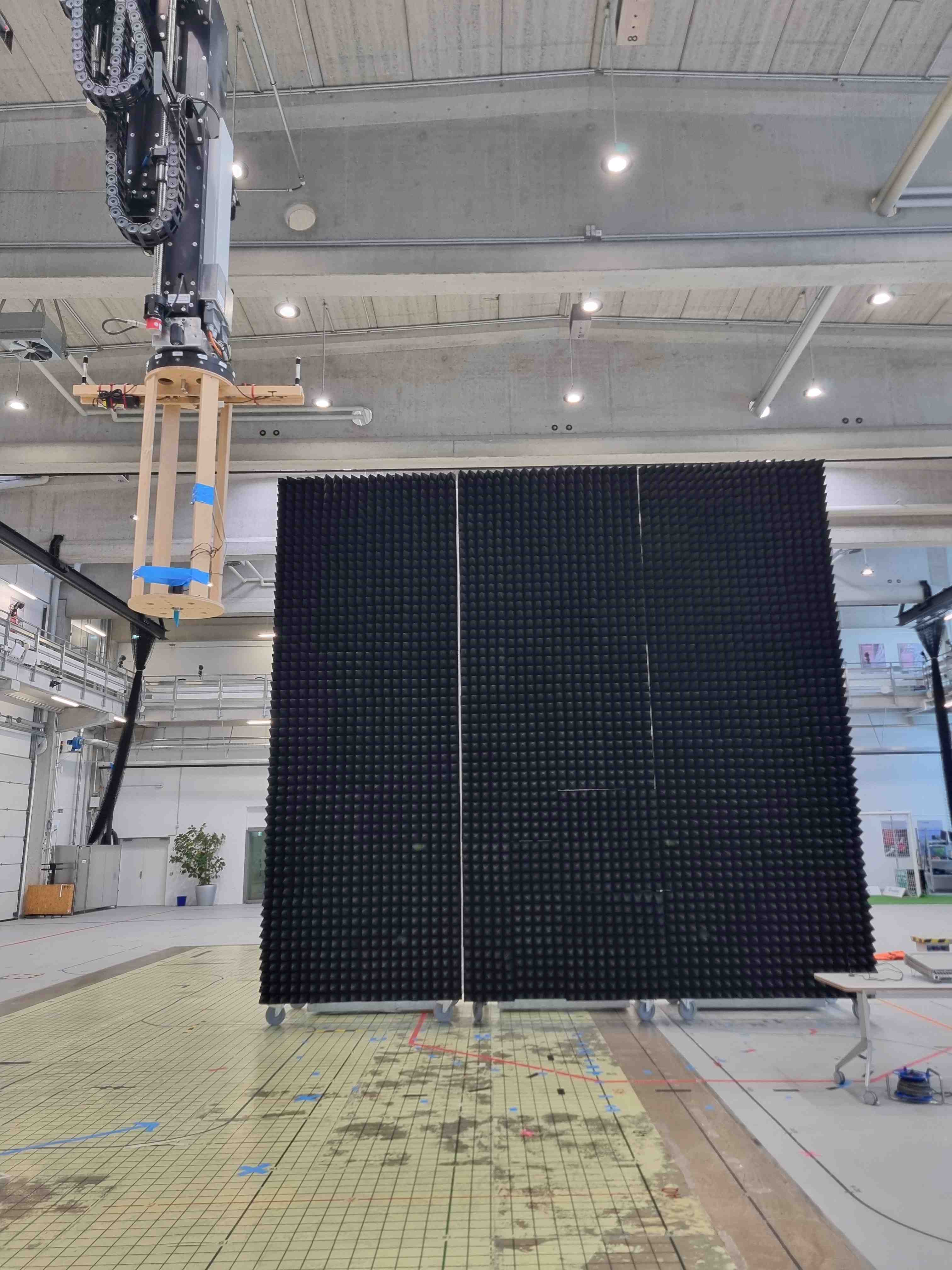}
        \subcaption{Wall 4.}
        \label{figure_dataset_pictures5}
    \end{minipage}
    \caption{Visualization of the recording setup, featuring the jammer (a) positioned on the 3D positioning system (e) and the arrangement of absorber walls around the antenna (b to e).}
    \label{figure_dataset_pictures}
\end{figure}

\begin{figure}[!t]
    \centering
	\begin{minipage}[t]{1.0\linewidth}
        \centering
    	\begin{minipage}[t]{0.425\linewidth}
            \centering
            \includegraphics[trim=10 10 10 10, clip, width=1.0\linewidth]{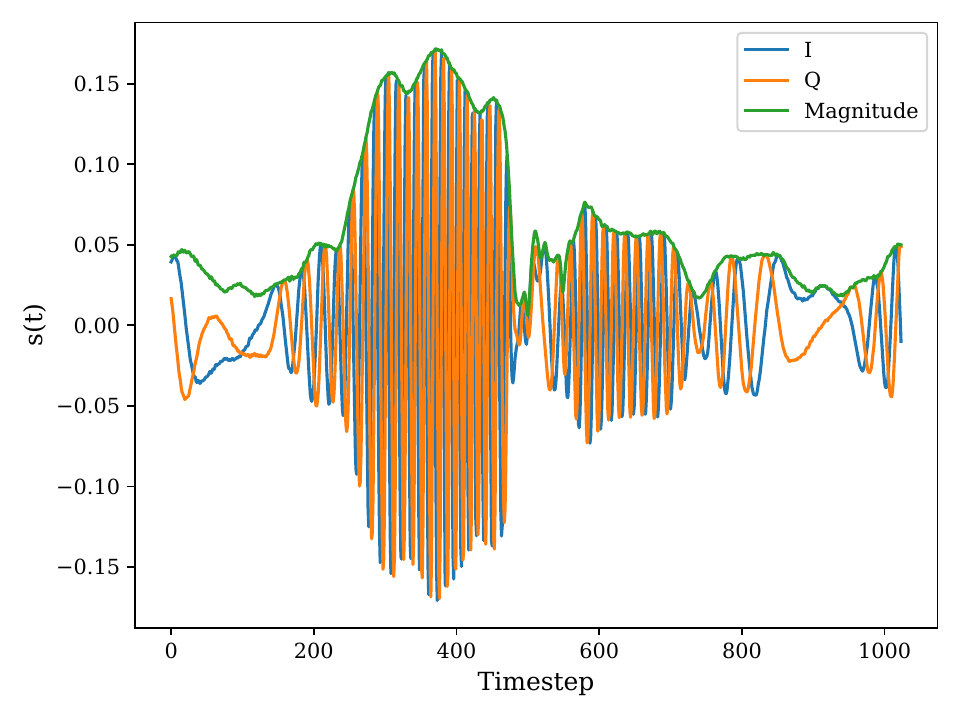}
        \end{minipage}
        \hfill
    	\begin{minipage}[t]{0.425\linewidth}
            \centering
            \includegraphics[trim=10 10 10 10, clip, width=1.0\linewidth]{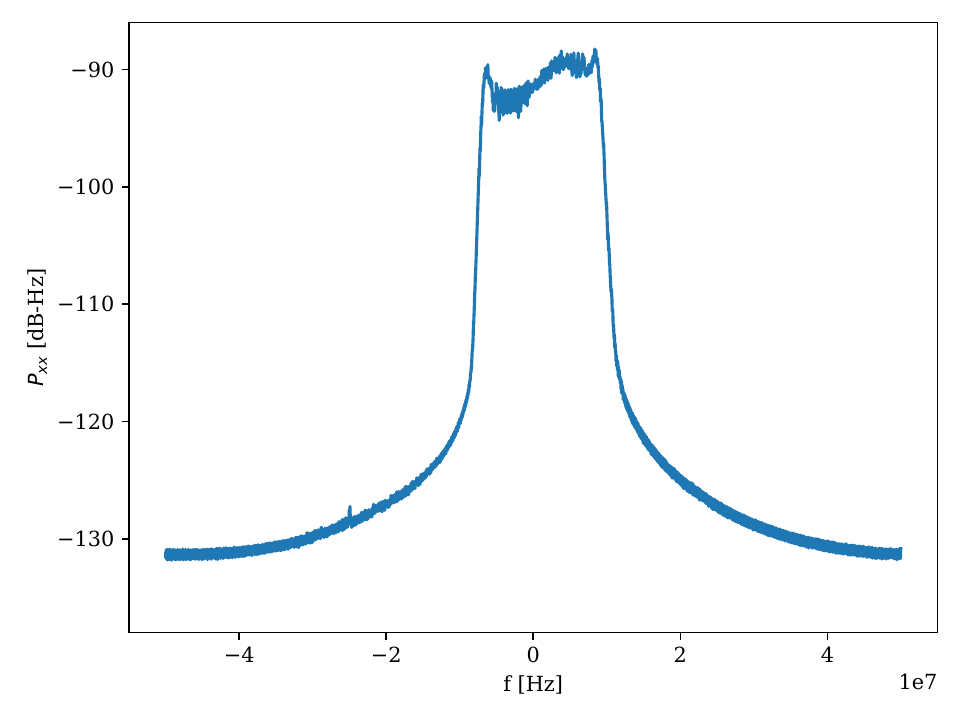}
        \end{minipage}
        \hfill
    	\begin{minipage}[t]{0.117\linewidth}
            \centering
            \includegraphics[trim=170 10 170 10, clip, width=1.0\linewidth]{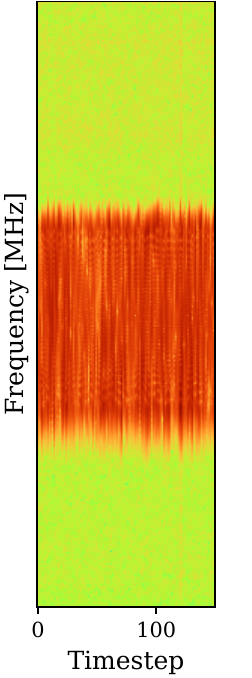}
        \end{minipage}
    \end{minipage}
    \hfill
	\begin{minipage}[t]{1.0\linewidth}
        \centering
    	\begin{minipage}[t]{0.425\linewidth}
            \centering
            \includegraphics[trim=10 10 10 10, clip, width=1.0\linewidth]{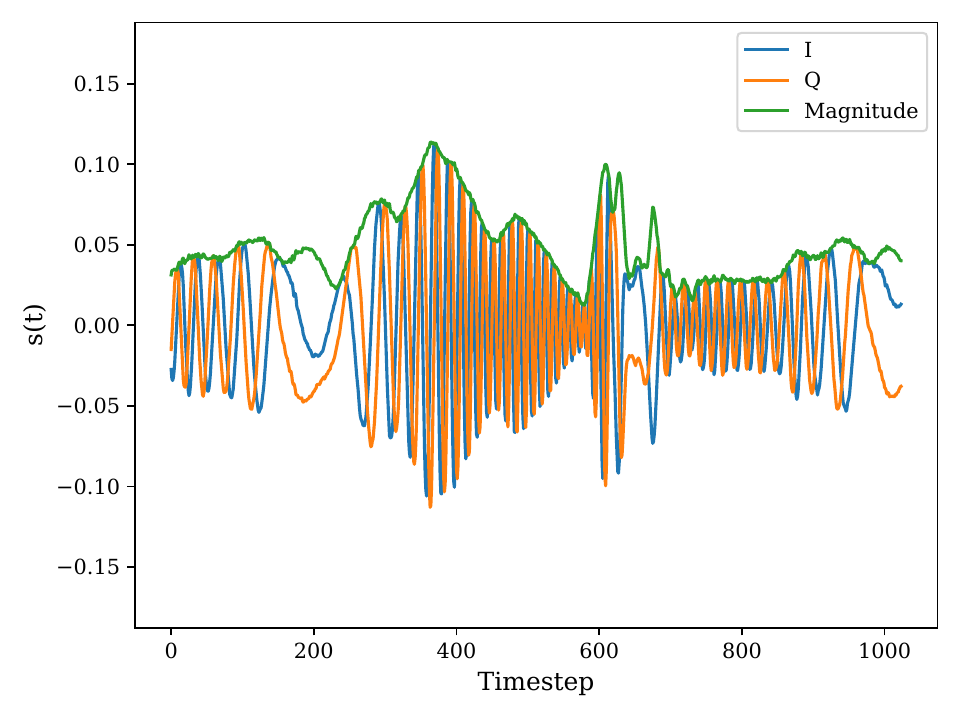}
        \end{minipage}
        \hfill
    	\begin{minipage}[t]{0.425\linewidth}
            \centering
            \includegraphics[trim=10 10 10 10, clip, width=1.0\linewidth]{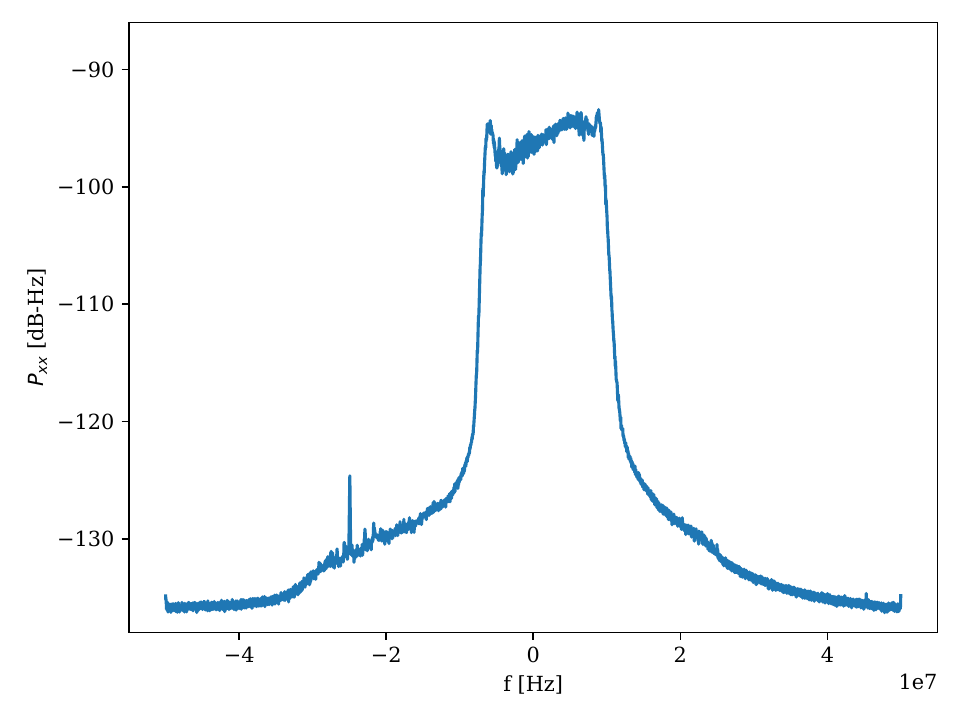}
        \end{minipage}
        \hfill
    	\begin{minipage}[t]{0.117\linewidth}
            \centering
            \includegraphics[trim=170 10 170 10, clip, width=1.0\linewidth]{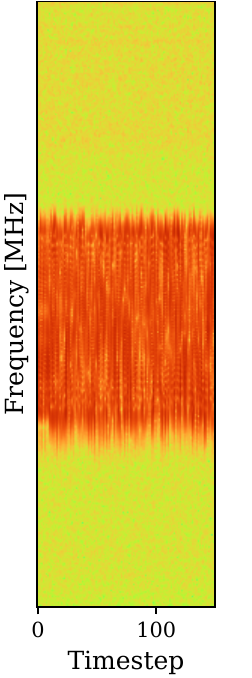}
        \end{minipage}
    \end{minipage}
    \caption{Visualization of IQ samples (left), Welch spectrum estimates (middle), and FFT-based spectrograms (right) without (top) and with (bottom) multipath effects.}
    \label{figure_dataset_iq_crpa}
\end{figure}

\newcommand\ac{0.119}
\begin{figure*}[!t]
    \centering
	\begin{minipage}[t]{\ac\linewidth}
        \centering
        \includegraphics[trim=8 10 10 10, clip, width=1.0\linewidth]{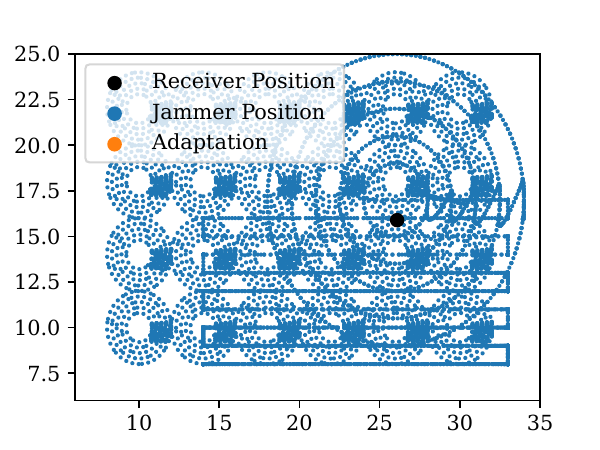}
        \subcaption{Random.}
        \label{figure_dataset_trajectories1}
    \end{minipage}
    \hfill
	\begin{minipage}[t]{\ac\linewidth}
        \centering
        \includegraphics[trim=8 10 10 10, clip, width=1.0\linewidth]{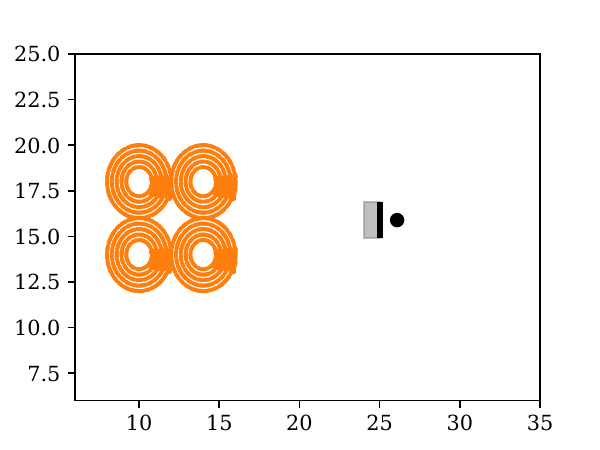}
        \subcaption{Wall 1.}
        \label{figure_dataset_trajectories2}
    \end{minipage}
    \hfill
	\begin{minipage}[t]{\ac\linewidth}
        \centering
        \includegraphics[trim=8 10 10 10, clip, width=1.0\linewidth]{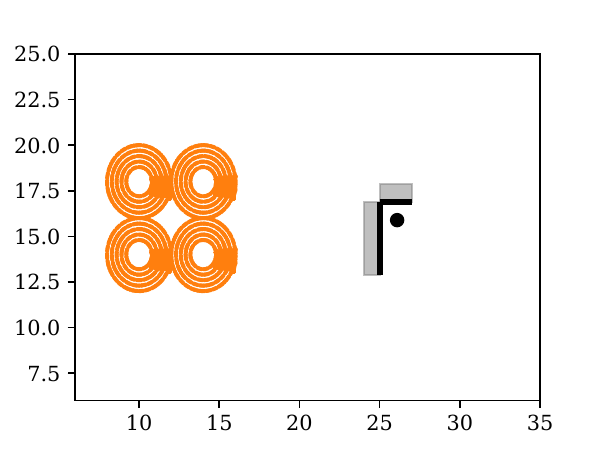}
        \subcaption{Wall 2.}
        \label{figure_dataset_trajectories3}
    \end{minipage}
    \hfill
	\begin{minipage}[t]{\ac\linewidth}
        \centering
        \includegraphics[trim=8 10 10 10, clip, width=1.0\linewidth]{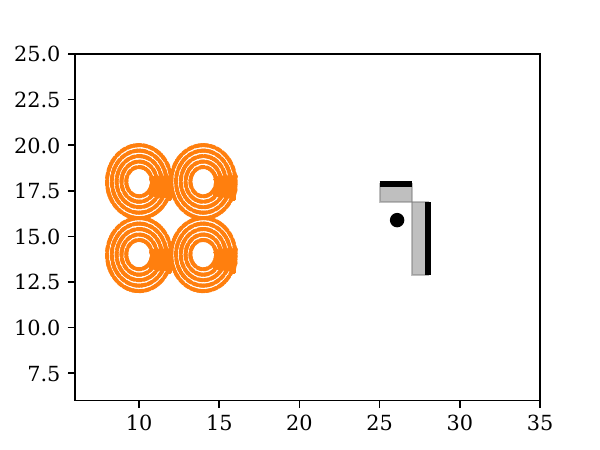}
        \subcaption{Wall 3.}
        \label{figure_dataset_trajectories4}
    \end{minipage}
    \hfill
	\begin{minipage}[t]{\ac\linewidth}
        \centering
        \includegraphics[trim=8 10 10 10, clip, width=1.0\linewidth]{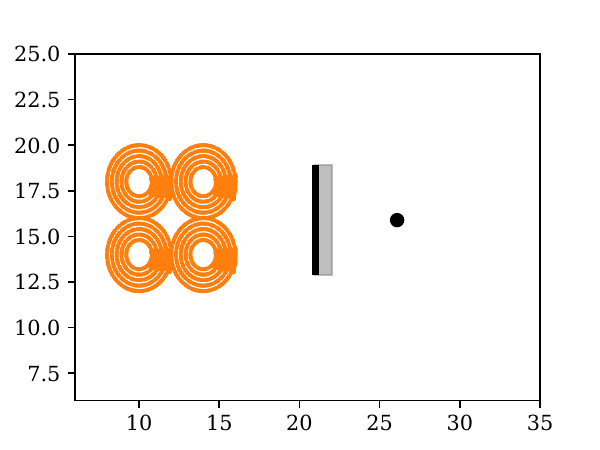}
        \subcaption{Wall 4.}
        \label{figure_dataset_trajectories5}
    \end{minipage}
    \hfill
	\begin{minipage}[t]{\ac\linewidth}
        \centering
        \includegraphics[trim=8 10 10 10, clip, width=1.0\linewidth]{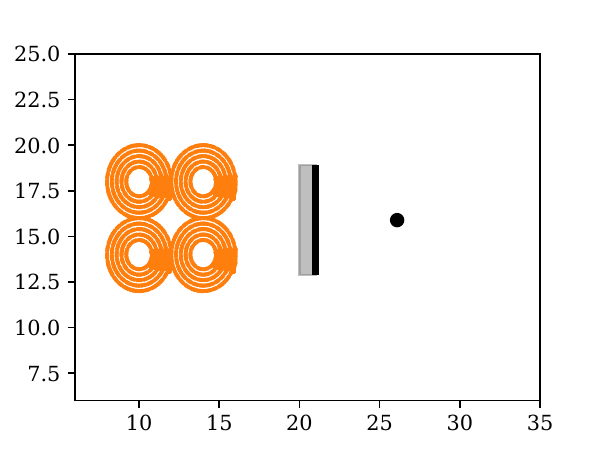}
        \subcaption{Wall 5.}
        \label{figure_dataset_trajectories6}
    \end{minipage}
    \hfill
	\begin{minipage}[t]{\ac\linewidth}
        \centering
        \includegraphics[trim=8 10 10 10, clip, width=1.0\linewidth]{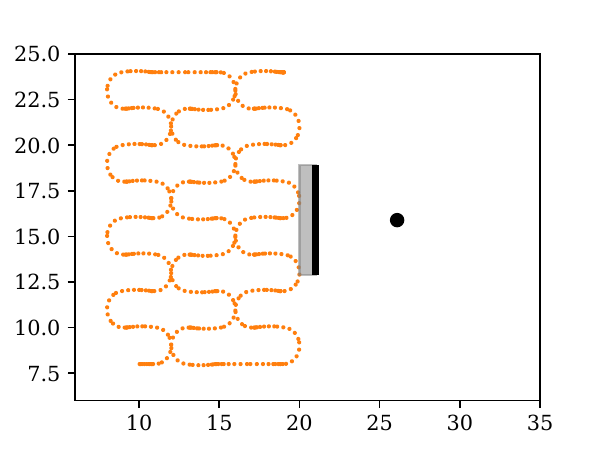}
        \subcaption{Meander.}
        \label{figure_dataset_trajectories7}
    \end{minipage}
    \hfill
	\begin{minipage}[t]{\ac\linewidth}
        \centering
        \includegraphics[trim=0 10 0 20, clip, width=1.0\linewidth]{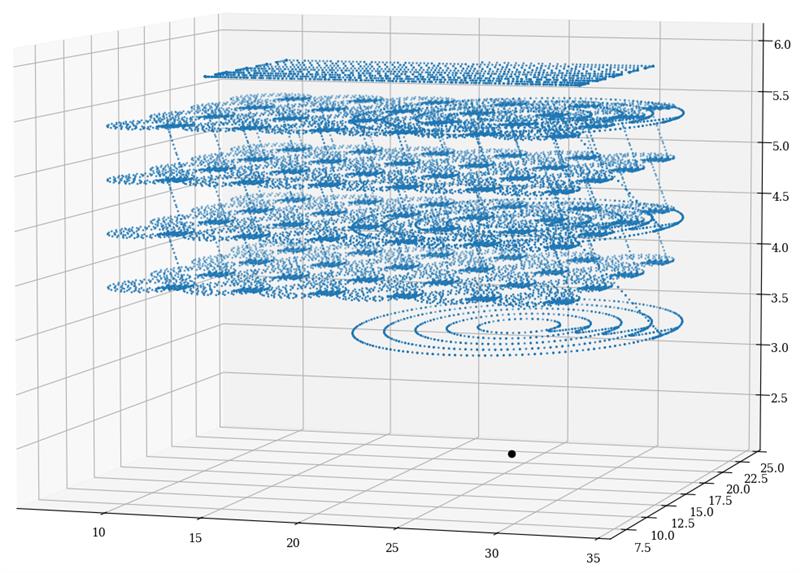}
        \subcaption{Height plot.}
        \label{figure_dataset_trajectories8}
    \end{minipage}
    \caption{The visualization depicts the trajectories (x in [m], y in [m]) of the positioning system under various conditions: random circular paths (a), differently positioned absorber walls (b to f), a meander trajectory (g), and varying heights (h). The black dot represents the antenna, with the absorber wall beside it and the black line indicates the absorbers' direction.}
    \label{figure_dataset_trajectories}
\end{figure*}

\begin{figure*}[!t]
    \centering
    \includegraphics[trim=10 10 10 10, clip, width=1.0\linewidth]{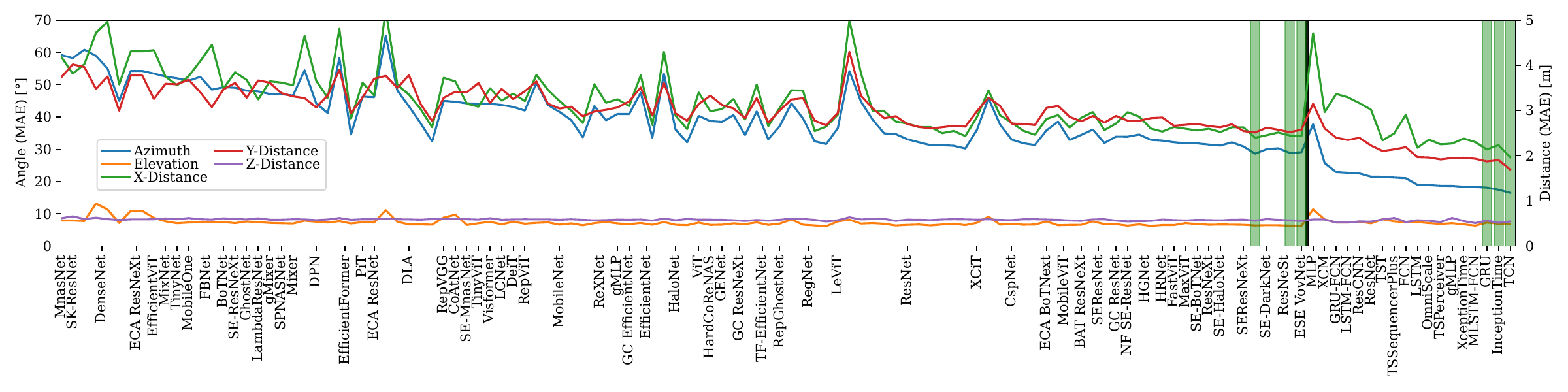}
    \vspace{-0.5cm}
    \caption{Evaluation results of 110 Hugging Face vision encoder models on spectrogram data (left of the black line) and 18 tsai models on IQ data (right of the black line). The top three models are highlighted in green (averaged over 10 trainings).}
    \label{figure_results_baselines}
\end{figure*}

\textbf{Recording Setup.} The primary objective of this study is to develop ML models for jammer localization that demonstrate robustness against various types of jammers, interference profiles, and environmental variations. To achieve this, we introduce a GNSS-based dataset collected at the Fraunhofer IIS L.I.N.K.~test and application center, a large indoor facility spanning 1,320$\,m^2$, which facilitates controlled data acquisition while accounting for multipath effects. The experimental setup consists of a receiver antenna, specifically a patch antenna array composed of four identical coaxial-fed square patch elements, positioned at one end of the hall. Data is recorded under two distinct scenarios: (1) a stationary MXG vector-signal generator placed at the opposite end of the hall, or (2) a mobile jammer device (Fig.~\ref{figure_dataset_pictures1}) mounted on a 3D positioning system (Fig.~\ref{figure_dataset_pictures5}), which enables precise recording of $x$-, $y$-, and $z$-coordinates with millimeter-level accuracy. In the first scenario, the MXG generator simulates common interference signals, including \textit{Chirp}, \textit{Frequency Hopping}, \textit{Modulated}, \textit{Multitone}, \textit{Pulsed}, and \textit{Noise}, with various subjammer types, spanning a bandwidth range of $0.2\,\text{MHz}$ to $60\,\text{MHz}$ and transmitting at power levels between $-20\,\text{dBm}$ and $10\,\text{dBm}$. In the second scenario, the mobile jammer transmits \textit{Chirp} signals with a bandwidth of $20\,\text{MHz}$. The received signals are captured as IQ snapshots, encompassing different interference conditions introduced by either the signal generator or the mobile jammer. A comprehensive dataset is collected under diverse conditions, including an unoccupied environment as well as configurations incorporating absorber walls. These walls are strategically positioned between the antenna and the generator, either with one or three absorber walls oriented towards or away from the antenna (Figures~\ref{figure_dataset_pictures2} to \ref{figure_dataset_pictures5}).

\textbf{Data Preprocessing.} Figure~\ref{figure_dataset_iq_crpa} presents exemplary data samples of a \textit{Chirp} interference signal, both in the absence and presence of multipath effects. The recorded data was acquired at a frequency of $1.57542\,\text{GHz}$ with a bandwidth of $100\,\text{MHz}$, quadrature sampled over a duration of $10\,\mu s$ (Figure~\ref{figure_dataset_iq_crpa}, left). The Welch plot (Figure~\ref{figure_dataset_iq_crpa}, middle) illustrates the power spectral density estimate of the signal, where the sharp increase and decrease in power delineate the bandwidth of the interfering signal. Non-overlapping spectrograms are generated using an FFT with a window size of 1,024 (Figure~\ref{figure_dataset_iq_crpa}, right). The primary objective of this analysis is to detect and accurately classify interference signals while localizing their source. The introduction of absorber walls results in reduced interference peaks in the IQ samples and lower signal power in the spectrograms. For data preprocessing, min-max normalization is applied to the spectrograms using normalization values of $-195.69$ and $-19.89$, while mean-standard deviation normalization is employed for IQ sample preprocessing, with separate normalization for each patch and IQ value.

\textbf{Dataset Overview.} Table~\ref{table_overview_dataset} provides a comprehensive overview of all recorded dataset subsets. Figure~\ref{figure_dataset_trajectories} illustrates the recorded trajectories of the mobile jammer. Initially, a primary train-test dataset was recorded across the entire area of the hall (Fig.~\ref{figure_dataset_trajectories1}), capturing five circular trajectories of varying diameters at four different heights $[3.9\,m, 4.4\,m, 4.9\,m, 5.4\,m]$, as depicted in Figure~\ref{figure_dataset_trajectories8}. This dataset comprises a total of 28,930 samples. Subsequently, additional scenarios were recorded, in which absorber walls were placed between the antenna and the mobile jammer. In these scenarios, data was collected for a $2 \times 2$ grid pattern, consisting of five circular trajectories at four different heights (Figures~\ref{figure_dataset_trajectories2} to \ref{figure_dataset_trajectories6}). Each recording contains approximately 3,600 samples. Finally, a dataset featuring a meandering trajectory was recorded (Fig.~\ref{figure_dataset_trajectories7}). The strategic placement of absorber walls facilitates the evaluation and benchmarking of the robustness of localization methods under varying environmental conditions, particularly in the presence of multipath effects.

\begin{table}[t!]
\begin{center}
    \caption{Overview of number of samples for all scenarios. $\rightarrow$ and $\leftarrow$ denotes the direction of the absorption.}
    \label{table_overview_dataset}
    \scriptsize \begin{tabular}{ p{0.5cm} | p{0.5cm} | p{0.5cm} | p{0.5cm} }
    \multicolumn{1}{c|}{\textbf{Dataset}} & \multicolumn{1}{c|}{\textbf{Set}} & \multicolumn{1}{c|}{\textbf{Samples}} & \multicolumn{1}{c}{\textbf{Comment}} \\ \hline
    \multicolumn{1}{c|}{Random} & \multicolumn{1}{c|}{Train} & \multicolumn{1}{r|}{23,140} & \multicolumn{1}{l}{No walls} \\
    \multicolumn{1}{c|}{Random} & \multicolumn{1}{c|}{Test} & \multicolumn{1}{r|}{5,790} & \multicolumn{1}{l}{No walls} \\
    \multicolumn{1}{c|}{Wall 1} & \multicolumn{1}{c|}{Adaptation / Test} & \multicolumn{1}{r|}{3,618} & \multicolumn{1}{l}{One wall, jammer $\rightarrow$ antenna} \\
    \multicolumn{1}{c|}{Wall 2} & \multicolumn{1}{c|}{Adaptation / Test} & \multicolumn{1}{r|}{3,617} & \multicolumn{1}{l}{Three walls, jammer $\rightarrow$ antenna} \\
    \multicolumn{1}{c|}{Wall 3} & \multicolumn{1}{c|}{Adaptation / Test} & \multicolumn{1}{r|}{3,625} & \multicolumn{1}{l}{Three walls, jammer $\leftarrow$ antenna} \\
    \multicolumn{1}{c|}{Wall 4} & \multicolumn{1}{c|}{Adaptation / Test} & \multicolumn{1}{r|}{3,627} & \multicolumn{1}{l}{One large wall, jammer $\leftarrow$ antenna} \\
    \multicolumn{1}{c|}{Wall 5} & \multicolumn{1}{c|}{Adaptation / Test} & \multicolumn{1}{r|}{3,628} & \multicolumn{1}{l}{One large wall, jammer $\rightarrow$ antenna} \\
    \multicolumn{1}{c|}{Meander} & \multicolumn{1}{c|}{Adaptation / Test} & \multicolumn{1}{r|}{884} & \multicolumn{1}{l}{One large wall, jammer $\rightarrow$ antenna} \\
    \end{tabular}
    \vspace{-0.3cm}
\end{center}
\end{table}

\section{Evaluation}
\label{label_evaluation}

\begin{table*}[t!]
\begin{center}
\setlength{\tabcolsep}{4.8pt}
    \caption{Overview of all jammer localization and interference classification errors. \textbf{Bold} denotes best results.}
    \label{table_all_results}
    \scriptsize \begin{tabular}{ p{0.5cm} | p{0.5cm} | p{0.5cm} | p{0.5cm} | p{0.5cm} | p{0.5cm} | p{0.5cm} | p{0.5cm} || p{0.5cm} | p{0.5cm} }
    & & \multicolumn{4}{c|}{\textbf{Distance error (MAE) [m]}} & \multicolumn{1}{c|}{\textbf{Azimuth}} & \multicolumn{1}{c||}{\textbf{Elevation}} & \multicolumn{2}{c}{\textbf{Classification [\%]}} \\
    \multicolumn{1}{c|}{\textbf{Method}} & \multicolumn{1}{c|}{\textbf{Data}} & \multicolumn{1}{c}{$\Delta x$} & \multicolumn{1}{c}{$\Delta y$} & \multicolumn{1}{c|}{$\Delta z$} & \multicolumn{1}{c|}{$\Delta d$} & \multicolumn{1}{c|}{\textbf{error} $\alpha$ [$^\circ$]} & \multicolumn{1}{c||}{\textbf{error} $\beta$ [$^\circ$]} & \multicolumn{1}{c|}{7 classes} & \multicolumn{1}{c}{310 subclasses} \\ \hline
    \multicolumn{1}{l|}{Hugging Face: VovNet~\cite{lee_park}} & \multicolumn{1}{l|}{FFT} & \multicolumn{1}{c}{2.403} & \multicolumn{1}{c}{2.564} & \multicolumn{1}{c|}{0.549} & \multicolumn{1}{c|}{3.557} & \multicolumn{1}{c|}{28.556} & \multicolumn{1}{c||}{6.114} & \multicolumn{1}{c|}{--} & \multicolumn{1}{c}{--} \\
    \multicolumn{1}{l|}{Hugging Face: SE-ResNeXt~\cite{xie_girshick}} & \multicolumn{1}{l|}{FFT} & \multicolumn{1}{c}{2.376} & \multicolumn{1}{c}{2.490} & \multicolumn{1}{c|}{0.552} & \multicolumn{1}{c|}{3.486} & \multicolumn{1}{c|}{27.991} & \multicolumn{1}{c||}{6.294} & \multicolumn{1}{c|}{--} & \multicolumn{1}{c}{--} \\
    \multicolumn{1}{l|}{Hugging Face: ResNeSt14d~\cite{zhang_wu_zhang}} & \multicolumn{1}{l|}{FFT} & \multicolumn{1}{c}{2.412} & \multicolumn{1}{c}{2.477} & \multicolumn{1}{c|}{0.564} & \multicolumn{1}{c|}{3.503} & \multicolumn{1}{c|}{28.253} & \multicolumn{1}{c||}{6.238} & \multicolumn{1}{c|}{--} & \multicolumn{1}{c}{--} \\
    \multicolumn{1}{l|}{tsai: TCN~\cite{bai_kolter_koltun}} & \multicolumn{1}{l|}{IQ} & \multicolumn{1}{c}{1.922} & \multicolumn{1}{c}{1.662} & \multicolumn{1}{c|}{0.547} & \multicolumn{1}{c|}{2.599} & \multicolumn{1}{c|}{16.150} & \multicolumn{1}{c||}{6.886} & \multicolumn{1}{c|}{--} & \multicolumn{1}{c}{--} \\
    \multicolumn{1}{l|}{tsai: InceptionTime~\cite{fawaz_lucas_forestier}} & \multicolumn{1}{l|}{IQ} & \multicolumn{1}{c}{2.234} & \multicolumn{1}{c}{1.869} & \multicolumn{1}{c|}{0.511} & \multicolumn{1}{c|}{2.957} & \multicolumn{1}{c|}{17.005} & \multicolumn{1}{c||}{6.880} & \multicolumn{1}{c|}{--} & \multicolumn{1}{c}{--} \\
    \multicolumn{1}{l|}{tsai: GRU~\cite{chung_gulcehre_cho}} & \multicolumn{1}{l|}{IQ} & \multicolumn{1}{c}{2.022} & \multicolumn{1}{c}{1.848} & \multicolumn{1}{c|}{0.562} & \multicolumn{1}{c|}{2.796} & \multicolumn{1}{c|}{17.545} & \multicolumn{1}{c||}{7.233} & \multicolumn{1}{c|}{--} & \multicolumn{1}{c}{--} \\
    \multicolumn{1}{l|}{ResNet18~\cite{he_zhang}, Quaternions} & \multicolumn{1}{l|}{FFT} & \multicolumn{1}{c}{--} & \multicolumn{1}{c}{--} & \multicolumn{1}{c|}{--} & \multicolumn{1}{c|}{--} & \multicolumn{1}{c|}{81.046} & \multicolumn{1}{c||}{9.455} & \multicolumn{1}{c|}{--} & \multicolumn{1}{c}{--} \\
    \multicolumn{1}{l|}{ResNet18~\cite{he_zhang}, Euler Angles} & \multicolumn{1}{l|}{FFT} & \multicolumn{1}{c}{4.588} & \multicolumn{1}{c}{3.412} & \multicolumn{1}{c|}{0.628} & \multicolumn{1}{c|}{5.752} & \multicolumn{1}{c|}{50.999} & \multicolumn{1}{c||}{9.353} & \multicolumn{1}{c|}{68.831} & \multicolumn{1}{c}{31.999} \\
    \multicolumn{1}{l|}{LSTM, Euler Angles} & \multicolumn{1}{l|}{IQ} & \multicolumn{1}{c}{2.116} & \multicolumn{1}{c}{1.877} & \multicolumn{1}{c|}{0.556} & \multicolumn{1}{c|}{2.882} & \multicolumn{1}{c|}{18.194} & \multicolumn{1}{c||}{6.850} & \multicolumn{1}{c|}{78.271} & \multicolumn{1}{c}{41.837} \\
    \multicolumn{1}{l|}{McAFF~\cite{zeng_gong_liu}} & \multicolumn{1}{l|}{IQ} & \multicolumn{1}{c}{2.081} & \multicolumn{1}{c}{1.771} & \multicolumn{1}{c|}{0.492} & \multicolumn{1}{c|}{2.776} & \multicolumn{1}{c|}{16.681} & \multicolumn{1}{c||}{6.097} & \multicolumn{1}{c|}{97.435} & \multicolumn{1}{c}{61.278} \\
    \multicolumn{1}{l|}{McAFF~\cite{zeng_gong_liu}} & \multicolumn{1}{l|}{FFT} & \multicolumn{1}{c}{2.999} & \multicolumn{1}{c}{2.544} & \multicolumn{1}{c|}{0.551} & \multicolumn{1}{c|}{3.972} & \multicolumn{1}{c|}{29.405} & \multicolumn{1}{c||}{7.517} & \multicolumn{1}{c|}{96.364} & \multicolumn{1}{c}{53.521} \\
    \multicolumn{1}{l|}{McAFF~\cite{zeng_gong_liu}} & \multicolumn{1}{l|}{CFO} & \multicolumn{1}{c}{2.172} & \multicolumn{1}{c}{1.957} & \multicolumn{1}{c|}{0.532} & \multicolumn{1}{c|}{2.972} & \multicolumn{1}{c|}{20.649} & \multicolumn{1}{c||}{7.084} & \multicolumn{1}{c|}{95.769} & \multicolumn{1}{c}{50.738} \\
    \multicolumn{1}{l|}{McAFF~\cite{zeng_gong_liu}} & \multicolumn{1}{l|}{STFT} & \multicolumn{1}{c}{2.620} & \multicolumn{1}{c}{2.419} & \multicolumn{1}{c|}{0.549} & \multicolumn{1}{c|}{3.608} & \multicolumn{1}{c|}{27.306} & \multicolumn{1}{c||}{7.153} & \multicolumn{1}{c|}{96.820} & \multicolumn{1}{c}{53.974} \\
    \multicolumn{1}{l|}{McAFF~\cite{zeng_gong_liu}} & \multicolumn{1}{l|}{IQ+CFO+STFT} & \multicolumn{1}{c}{1.975} & \multicolumn{1}{c}{1.733} & \multicolumn{1}{c|}{0.489} & \multicolumn{1}{c|}{2.672} & \multicolumn{1}{c|}{15.690} & \multicolumn{1}{c||}{6.072} & \multicolumn{1}{c|}{97.394} & \multicolumn{1}{c}{63.587} \\
    \multicolumn{1}{l|}{McAFF~\cite{zeng_gong_liu}} & \multicolumn{1}{l|}{IQ+FFT+CFO+STFT} & \multicolumn{1}{c}{1.980} & \multicolumn{1}{c}{1.731} & \multicolumn{1}{c|}{\textbf{0.488}} & \multicolumn{1}{c|}{2.675} & \multicolumn{1}{c|}{16.006} & \multicolumn{1}{c||}{6.102} & \multicolumn{1}{c|}{\textbf{97.508}} & \multicolumn{1}{c}{\textbf{63.753}} \\
    \multicolumn{1}{l|}{Fusion (ours), $\gamma=1.0$, first $\text{dropout}=50\%$} & \multicolumn{1}{l|}{IQ+FFT} & \multicolumn{1}{c}{1.718} & \multicolumn{1}{c}{1.698} & \multicolumn{1}{c|}{0.601} & \multicolumn{1}{c|}{2.489} & \multicolumn{1}{c|}{15.377} & \multicolumn{1}{c||}{6.459} & \multicolumn{1}{c|}{--} & \multicolumn{1}{c}{--} \\
    \multicolumn{1}{l|}{Fusion (ours), $\gamma=1.0$, first $\text{dropout}=50\%$} & \multicolumn{1}{l|}{IQ+FFT+AoA} & \multicolumn{1}{c}{\textbf{1.456}} & \multicolumn{1}{c}{\textbf{1.469}} & \multicolumn{1}{c|}{0.525} & \multicolumn{1}{c|}{\textbf{2.134}} & \multicolumn{1}{c|}{\textbf{13.725}} & \multicolumn{1}{c||}{\textbf{5.882}} & \multicolumn{1}{c|}{--} & \multicolumn{1}{c}{--} \\
    \end{tabular}
    \vspace{-0.3cm}
\end{center}
\end{table*}

\textbf{Encoder Models.} The same hardware setup\footnote{For all experiments, we use Nvidia Tesla V100-SXM2 GPUs with 32 GB VRAM equipped with Core Xeon CPUs and 192\,GB RAM. We use the SGD optimizer with a multi-step learning rate of $10^{-2}$, decay of $5 \cdot 10^{-4}$, momentum of 0.9, a batch size of 64, and train for 200 epochs. We train each model 10 times and present the mean accuracy for classification tasks, as well as the mean absolute error (MAE) for regression tasks in $m$ and $^\circ$.} is utilized for all experiments. Figure~\ref{figure_results_baselines} presents an evaluation of all IQ and FFT-based encoder models. The lowest error on FFT-computed data is achieved by VovNet~\cite{lee_park}, SE-ResNeXT~\cite{xie_girshick}, and ResNeSt14d~\cite{zhang_wu_zhang}, which are highlighted in green. However, the azimuth estimation remains relatively high at $27.991^\circ$. In contrast, for IQ samples, the time-series models TCN~\cite{bai_kolter_koltun}, InceptionTime~\cite{fawaz_lucas_forestier}, and GRU~\cite{chung_gulcehre_cho} achieve a significantly lower azimuth error of $16.15^\circ$. These findings motivate the incorporation of VoVNet and TCN into our fusion architecture.

\begin{figure}[!t]
    \centering
	\begin{minipage}[t]{0.492\linewidth}
        \centering
        \includegraphics[trim=9 9 9 9, clip, width=1.0\linewidth]{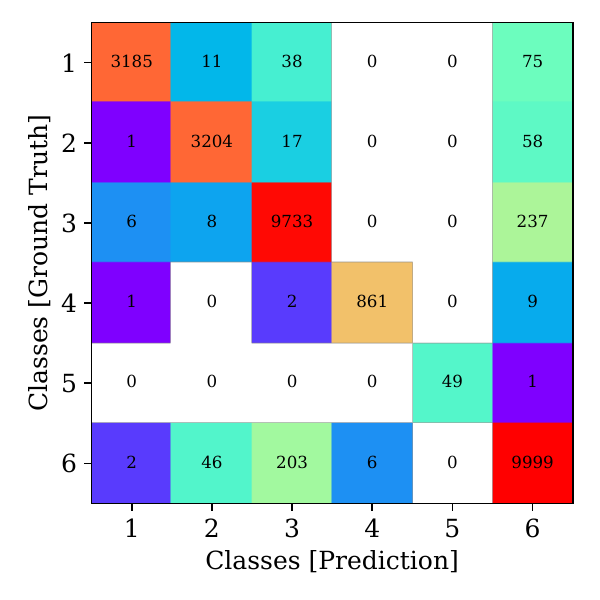}
        \subcaption{6 interference classes.}
        \label{figure_confusion_matrices1}
    \end{minipage}
    \hfill
	\begin{minipage}[t]{0.492\linewidth}
        \centering
        \includegraphics[trim=7 9 9 9, clip, width=1.0\linewidth]{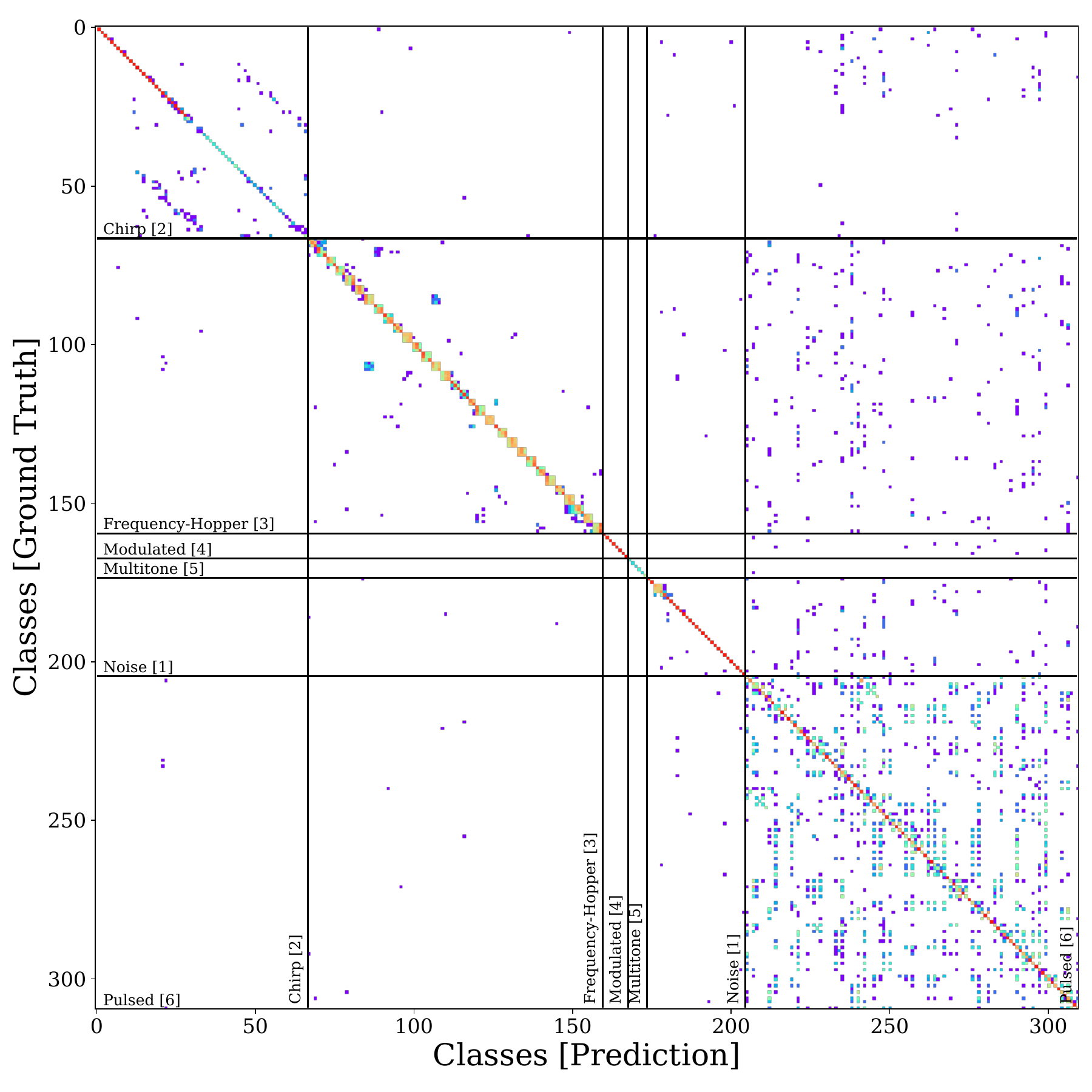}
        \subcaption{310 interference subclasses.}
        \label{figure_confusion_matrices2}
    \end{minipage}
    \caption{Confusion matrices for McAFF~\cite{zeng_gong_liu}.}
    \label{figure_confusion_matrices}
\end{figure}

\textbf{Method Evaluation.} Next, we compare the results of all methods presented in Table~\ref{table_all_results}. In addition to vision encoders and time-series models, we train ResNet18~\cite{he_zhang} (with an $\alpha$ error of $50.999^\circ$) and an LSTM (with an $\alpha$ error of $18.194^\circ$), as these models are well-established for GNSS interference characterization~\cite{brieger_ion_gnss,ott_heublein_icl,heublein_feigl_posnav,heublein_feigl_crpa}. Consequently, a specialized ResNet model is required for the localization task. Furthermore, we assess different loss functions for direction estimation, employing either Quaternions or Euler angles. Since Euler angles demonstrate significantly superior performance, we adopt this approach for all subsequent training procedures. We evaluate McAFF in six distinct configurations, utilizing the IQ, FFT, CFO, and STFT pipelines either individually or in combination. While the IQ-only approach yields the best results ($\alpha = 16.681^\circ$, $\beta = 6.097^\circ$), significantly outperforming the FFT-only, CFO-only, and STFT-only approaches, further improvements are achieved by combining IQ, CFO, and STFT ($\alpha = 15.69^\circ$, $\beta = 6.072^\circ$). Notably, the FFT component only influences the classification task. Our fusion method achieves the lowest errors, with $\Delta d = 2.134\,m$, $\alpha = 13.725^\circ$, and $\beta = 5.882^\circ$, by incorporating the 22 statistical AoA features. This result underscores the significance of integrating multiple data representations for improved performance. Figure~\ref{figure_confusion_matrices} provides a detailed analysis of the classification task. The most frequent misclassifications occur in the \textit{FrequHopper} (class 3) and \textit{Pulsed} (class 6) categories. However, the model successfully classifies all 310 subclasses, with the highest degree of confusion observed within the \textit{Pulsed} subclass.

\begin{figure}[!t]
    \centering
    \hfill
	\begin{minipage}[t]{0.45\linewidth}
        \centering
        \includegraphics[trim=10 10 10 10, clip, width=1.0\linewidth]{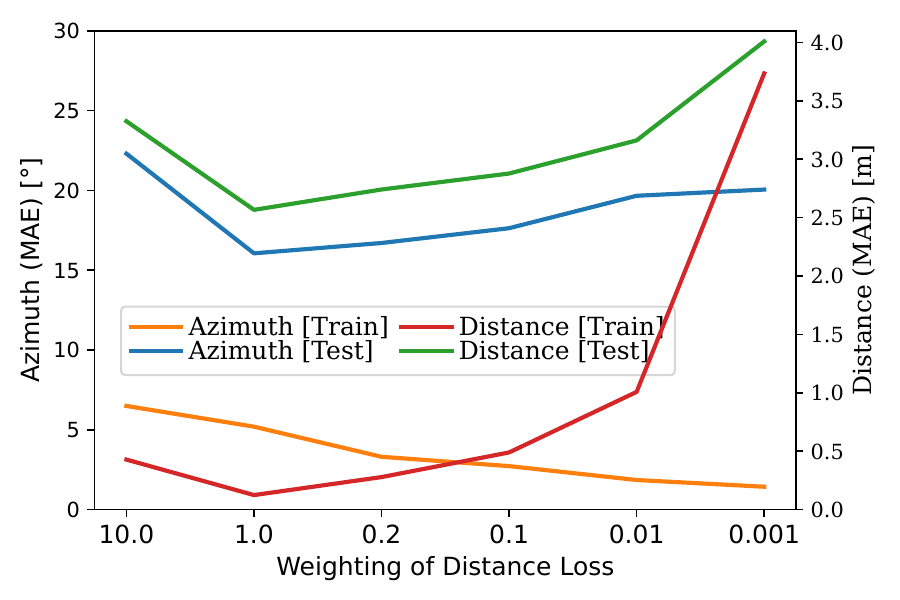}
        \caption{Weighting of distance loss function.}
        \label{figure_hyperparameter_weighting}
    \end{minipage}
    \hfill
	\begin{minipage}[t]{0.45\linewidth}
        \centering
        \includegraphics[trim=10 10 10 10, clip, width=1.0\linewidth]{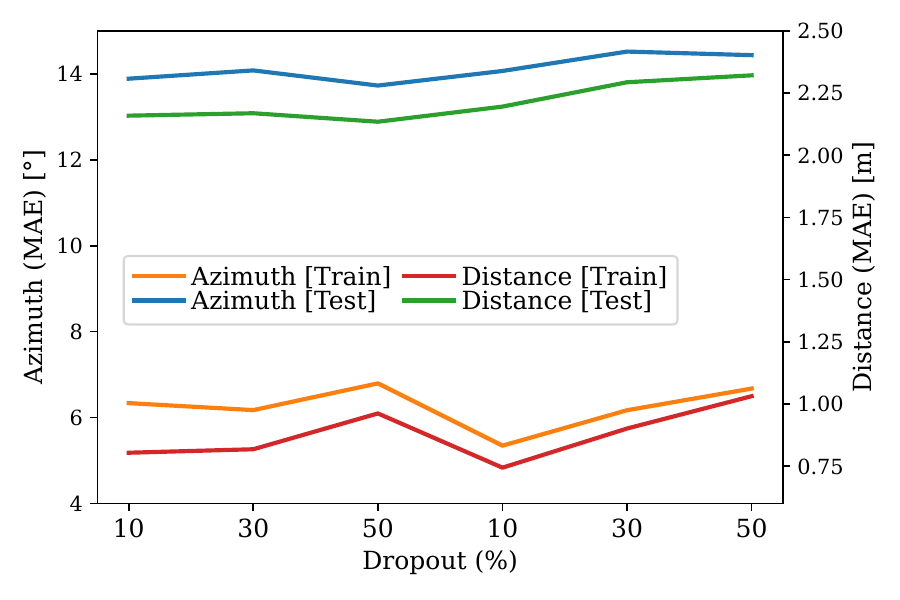}
        \caption{Hyperparameter search for dropout rate.}
        \label{figure_hyperparameter_dropout}
    \end{minipage}
    \hfill
\end{figure}

\textbf{Hyperparameter Selection.} Figure~\ref{figure_hyperparameter_weighting} presents the results for the weighting of the distance loss function, with the lowest errors observed at $\gamma = 1.0$, which is selected for subsequent training. However, the discrepancy between the training and test datasets indicates model overfitting. To mitigate this, we investigate the optimal dropout rate (Figure~\ref{figure_hyperparameter_dropout}), identifying a dropout rate of 50\% applied exclusively before concatenation as the most effective configuration.

\newcommand\ad{0.129}
\begin{figure*}[!t]
    \centering
	\begin{minipage}[t]{0.19\linewidth}
        \centering
        \includegraphics[trim=10 10 30 10, clip, width=1.0\linewidth]{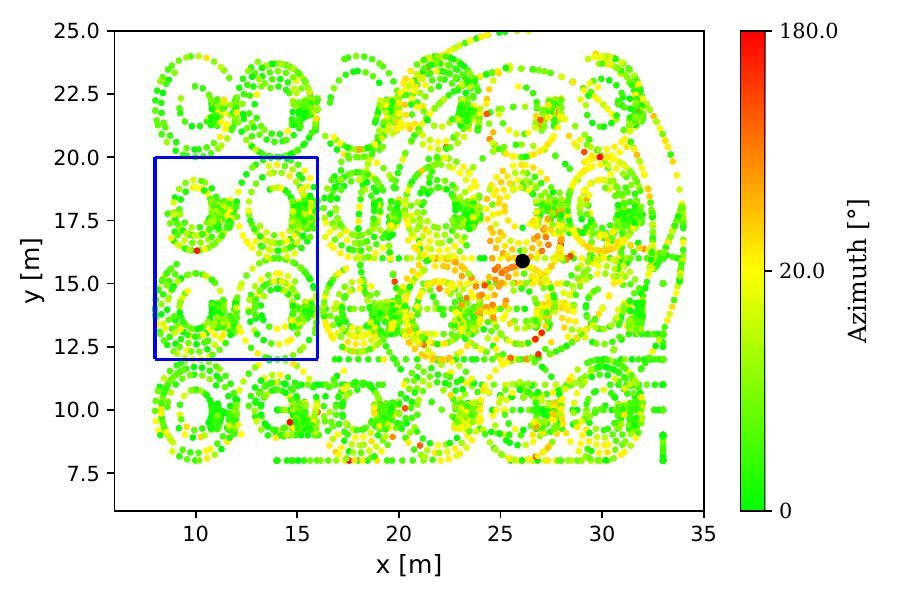}
        \subcaption{Random.}
        \label{figure_trajectory_predictions1}
    \end{minipage}
    \hfill
	\begin{minipage}[t]{\ad\linewidth}
        \centering
        \includegraphics[trim=10 10 10 10, clip, width=1.0\linewidth]{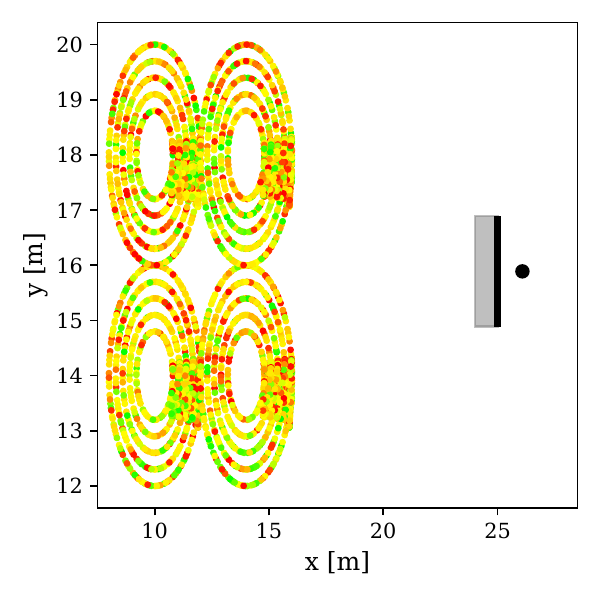}
        \subcaption{Wall 1.}
        \label{figure_trajectory_predictions2}
    \end{minipage}
    \hfill
	\begin{minipage}[t]{\ad\linewidth}
        \centering
        \includegraphics[trim=10 10 10 10, clip, width=1.0\linewidth]{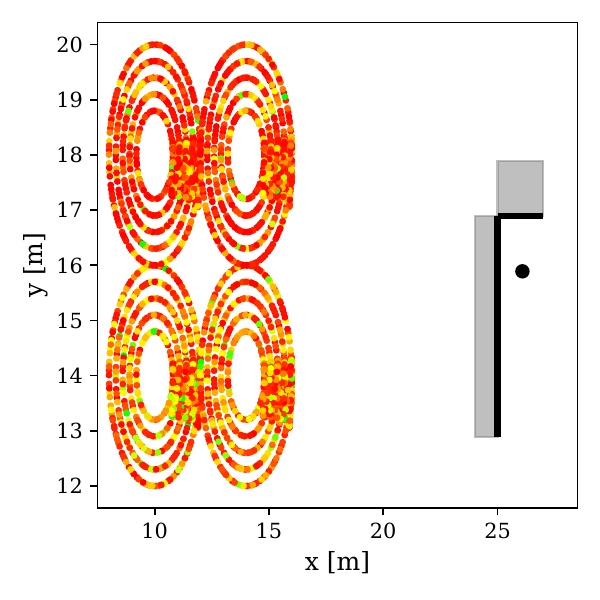}
        \subcaption{Wall 2.}
        \label{figure_trajectory_predictions3}
    \end{minipage}
    \hfill
	\begin{minipage}[t]{\ad\linewidth}
        \centering
        \includegraphics[trim=10 10 10 10, clip, width=1.0\linewidth]{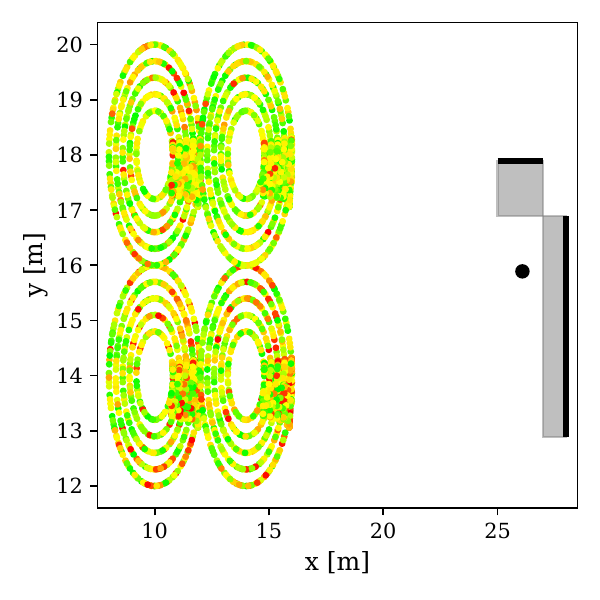}
        \subcaption{Wall 3.}
        \label{figure_trajectory_predictions4}
    \end{minipage}
    \hfill
	\begin{minipage}[t]{\ad\linewidth}
        \centering
        \includegraphics[trim=10 10 10 10, clip, width=1.0\linewidth]{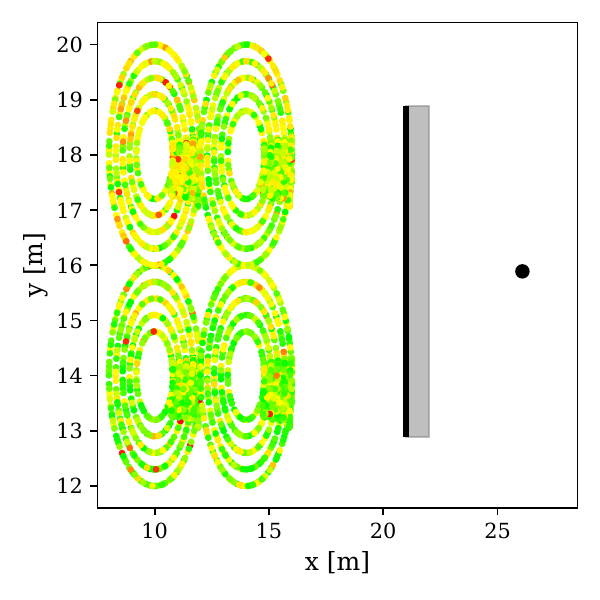}
        \subcaption{Wall 4.}
        \label{figure_trajectory_predictions5}
    \end{minipage}
    \hfill
	\begin{minipage}[t]{\ad\linewidth}
        \centering
        \includegraphics[trim=10 10 10 10, clip, width=1.0\linewidth]{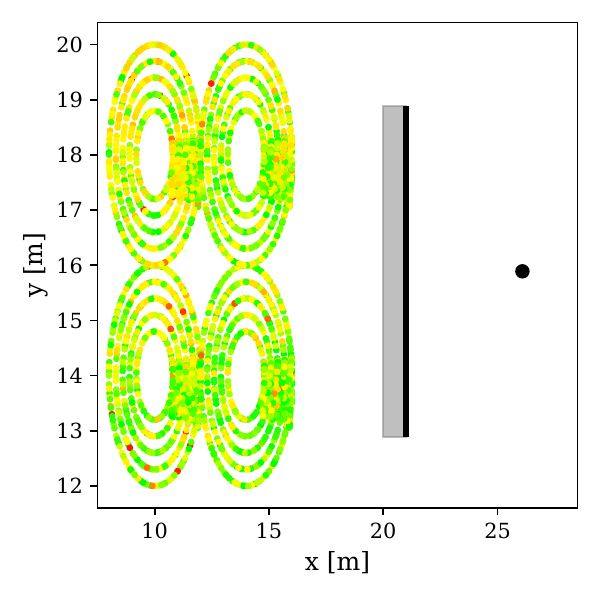}
        \subcaption{Wall 5.}
        \label{figure_trajectory_predictions6}
    \end{minipage}
    \hfill
	\begin{minipage}[t]{\ad\linewidth}
        \centering
        \includegraphics[trim=10 10 10 10, clip, width=1.0\linewidth]{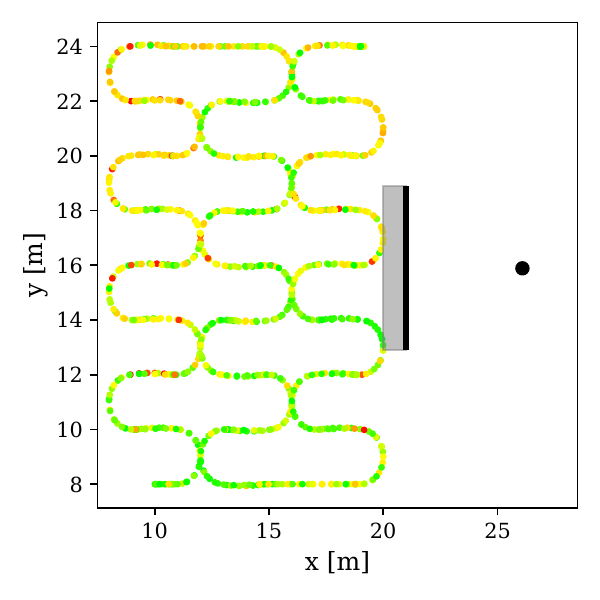}
        \subcaption{Meander.}
        \label{figure_trajectory_predictions7}
    \end{minipage}
    \caption{Evaluation of azimuth prediction, based on the position, for our fusion method across all test datasets. The blue rectangle highlights the region of the $2 \times 2$ grid of circles with absorber walls. The results correspond to Table~\ref{table_cross_validation}.}
    \label{figure_trajectory_predictions}
    \vspace{-0.25cm}
\end{figure*}

\begin{table}[t!]
\begin{center}
    \caption{Overview of results for all evaluation datasets.}
    \label{table_cross_validation}
    \scriptsize \begin{tabular}{ p{0.5cm} | p{0.5cm} | p{0.5cm} | p{0.5cm} | p{0.5cm} | p{0.5cm} | p{0.5cm} | p{0.5cm} }
    & \multicolumn{4}{c|}{\textbf{Distance error (MAE) [m]}} & \multicolumn{1}{c|}{\textbf{Azimuth}} & \multicolumn{1}{c}{\textbf{Elevation}} \\
    \multicolumn{1}{c|}{\textbf{Test}} & \multicolumn{1}{c}{$\Delta x$} & \multicolumn{1}{c}{$\Delta y$} & \multicolumn{1}{c|}{$\Delta z$} & \multicolumn{1}{c|}{$\Delta d$} & \multicolumn{1}{c|}{\textbf{error} $\alpha$ [$^\circ$]} & \multicolumn{1}{c}{\textbf{error} $\beta$ [$^\circ$]} \\ \hline
    \multicolumn{1}{c|}{Random} & \multicolumn{1}{r}{1.456} & \multicolumn{1}{r}{1.469} & \multicolumn{1}{r|}{0.525} & \multicolumn{1}{r|}{2.134} & \multicolumn{1}{r|}{13.725} & \multicolumn{1}{c}{5.882} \\
    \multicolumn{1}{c|}{Wall 1} & \multicolumn{1}{r}{6.002} & \multicolumn{1}{r}{3.383} & \multicolumn{1}{r|}{0.494} & \multicolumn{1}{r|}{6.907} & \multicolumn{1}{r|}{49.702} & \multicolumn{1}{c}{4.037} \\
    \multicolumn{1}{c|}{Wall 2} & \multicolumn{1}{r}{12.587} & \multicolumn{1}{r}{3.432} & \multicolumn{1}{r|}{0.490} & \multicolumn{1}{r|}{13.055} & \multicolumn{1}{r|}{115.804} & \multicolumn{1}{c}{5.059} \\
    \multicolumn{1}{c|}{Wall 3} & \multicolumn{1}{r}{7.085} & \multicolumn{1}{r}{2.587} & \multicolumn{1}{r|}{0.506} & \multicolumn{1}{r|}{7.559} & \multicolumn{1}{r|}{32.222} & \multicolumn{1}{c}{7.174} \\
    \multicolumn{1}{c|}{Wall 4} & \multicolumn{1}{r}{2.492} & \multicolumn{1}{r}{3.181} & \multicolumn{1}{r|}{0.509} & \multicolumn{1}{r|}{4.073} & \multicolumn{1}{r|}{17.418} & \multicolumn{1}{c}{2.803} \\
    \multicolumn{1}{c|}{Wall 5} & \multicolumn{1}{r}{2.733} & \multicolumn{1}{r}{3.331} & \multicolumn{1}{r|}{0.513} & \multicolumn{1}{r|}{4.339} & \multicolumn{1}{r|}{19.033} & \multicolumn{1}{c}{2.912} \\
    \multicolumn{1}{c|}{Meander} & \multicolumn{1}{r}{4.110} & \multicolumn{1}{r}{3.736} & \multicolumn{1}{r|}{0.893} & \multicolumn{1}{r|}{5.625} & \multicolumn{1}{r|}{23.295} & \multicolumn{1}{c}{4.170} \\
    \end{tabular}
\end{center}
\end{table}

\textbf{Robustness to NLoS Changes.} Furthermore, we assess the robustness of the proposed method against environmental variations, particularly changes in multipath effects caused by NLoS scenarios through the placement of absorber walls. The model is trained on the complete random dataset under LoS conditions and tested across all NLoS scenarios. A summary of the evaluation results is provided in Table~\ref{table_cross_validation}, while Figure~\ref{figure_trajectory_predictions} illustrates the azimuth error as a function of the reference position. For the LoS scenario, the azimuth error remains low ($\alpha = 13.725^\circ$), with a slight increase at shorter antenna distances (Figure~\ref{figure_trajectory_predictions1}). However, the error significantly increases in the presence of absorber walls. Notably, the model maintains robustness when the absorber is positioned directly in front of the antenna or when the walls are widely spaced ($\alpha = 17.418^\circ$), as shown in Figures~\ref{figure_trajectory_predictions5} and \ref{figure_trajectory_predictions6}. In contrast, the error increases substantially in scenario 2 (Figure~\ref{figure_trajectory_predictions3}). These findings highlight the potential of domain-incremental learning as a promising direction for future research.
\section{Conclusion}
\label{label_conclusion}

We collected an indoor dataset for GNSS jammer localization and proposed a fusion method integrating IQ, FFT, and statistical AoA features. Our approach achieved a distance error of $2.134\,m$, an azimuth error of $13.725^\circ$, and an elevation error of $5.882^\circ$, surpassing the performance of classical baseline methods and novel ML-based techniques such as McAFF.

\section*{Acknowledgments}
\small This work has been carried out within the PaiL project, funding code 50NP2506, and DARCII project, funding code 50NA2401, sponsored by the German Federal Ministry for Economic Affairs and Climate Action (BMWK) and supported by the German Aerospace Center (DLR), the Bundesnetzagentur (BNetzA), and the Federal Agency for Cartography and Geodesy (BKG).

\bibliography{RadarConf2025}
\bibliographystyle{IEEEtran}

\end{document}